\documentclass[numreferences]{kluwer}
\newtheorem{theorem}{Theorem}
\newcommand{\submitted}{Submitted}

\newcommand{\J}{\ensuremath{\mathcal{J}}}
\renewcommand{\S}{\ensuremath{\mathcal{S}}}
\newcommand{\di}[2]{{}_{#1}^{({#2})}}
\newcommand{\maple}{{\tt MAPLE V}\texttrademark}
\providecommand{\texttrademark}{${}^{\mbox{\scriptsize TM}}$}

\begin{document}
\begin{opening}
\title{
On the extrapolation of perturbation series\thanks{Talk
presented at the {\it International Conference on Rational
Approximation} (ICRA99), Antwerpen, Belgium, 1999}
}
\date{28/01/2000, 12:14}
\author{Herbert H.\ H.\
\surname{Homeier}\thanks{Homepage:
http://homepages.uni-regensburg.de/$\sim$hoh05008}}
\institute{Institut f\"ur Physikalische und Theoretische Chemie,
Universit\"at Regensburg,\\
D-93040 Regensburg,
Germany\\
\email{Herbert.Homeier@na-net.ornl.gov}}
\runningauthor{H. H. H. Homeier}
\runningtitle{Extrapolation of Perturbation Series}
\begin{ao}
Priv.-Doz. Dr. H. H. H. Homeier\\
Institut f\"ur Physikalische und Theoretische Chemie\\
Universit\"at Regensburg\\
Universit\"atsstr.\ 31\\
D-93053 Regensburg\\
Germany\\
e-mail: Herbert.Homeier@na-net.ornl.gov
\end{ao}
%
\begin{abstract}
We discuss certain special cases of algebraic approximants that
are given as zeroes of so-called {\em effective characteristic
polynomials} and their generalization to a multiseries setting.
These approximants are useful for the convergence acceleration
or summation of quantum mechanical perturbation series. Examples
will be given and some properties will be discussed.
\end{abstract}
\classification{Mathematics Subject Classification (1991)}{Primary 65B05; Secondary 65B10 40A05 40A25}
\end{opening}
\section{Introduction}
    In quantum mechanics, the usual Rayleigh-Schr\"odinger Perturbation
    Theory (RSPT) is used for solving the eigenvalue problem of a
    Hamilton operator
    \begin{equation}
    H = H_0 + \beta V
    \end{equation}
    in terms of an unperturbed Hamiltonian $H_0$ with known spectrum, a
    perturbation $V$ and a coupling constant $\beta>0$. The RSPT yields
    for the $I$-th  eigenvalue a formal power series
    \begin{equation}\label{eqEIbeta}
    E^{(I)}(\beta) = E^{(I)}_0 + \beta E^{(I)}_1 + \dots +
    \beta^n E^{(I)}_n+ \dots
    \end{equation}
    with real coefficients $E^{(I)}_n$.
    Such series are often divergent, for instance for the {\it quartic
    anharmonic oscillator} (AHO) with Hamiltonian
    \begin{equation}\label{eqquarticAHO}
    \displaystyle{H} = -\frac{d^{2}}{dx^2} + {x}^{2}  +
    \beta {x}^{4}
    \end{equation}
    on the real line. For this and closely related problems, there is a vast amount of
    literature
    since the
    quartic AHO may be considered as the prototype of a
    zero-dimensional $\Phi^4$ field theory, and thus, it has been used
    as test case for almost every important numerical method to solve
    the quantum mechanical eigenvalue problem. A recent example is the
    use of the generalized Bloch equation as an iterative method for
    the solution of the Schr\"odinger equation
    \cite{Meissner95,MeissnerSteinborn97ani,MeissnerSteinborn97qsa}.
    For the quartic AHO, the $n$-th term $a_n$ in the RSPT series
    behaves as $a_n \sim
    -\sqrt{24/\pi^3}(-3\beta/2)^{n}\Gamma(n+1/2)$ for
    large~$n$ \cite{BenderWu69,BenderWu71,BenderWu73} and the series
    has zero radius of convergence in the variable $\beta$. Thus, one
    has to sum such alternating divergent series.

    There are many summation methods that can be used in principle, a
    review of which is outside the scope of the present article. 
    Consider a (formal) power series
    \begin{equation}
    f(\beta) \; = \; \sum_{j=0}^{\infty} \> c_{j} \, \beta^{j}\>
    \end{equation}
    with partial sums
    \begin{equation}
    s_n \; = \; \sum_{j=0}^{n} \> c_{j} \, \beta^{j}\>.
    \end{equation}
    An
    important
    class of summation methods are nonlinear sequence transformations that
    transform the sequence $\{s_n\}_{n=0}^{\infty}$ of partial sums
    to a new transformed sequence $\{s_n'\}_{n=0}^{\infty}$ that is assumed
    to converge to the so-called antilimit of the divergent sequence
    $\{s_n\}$. This antilimit is taken as the result of the summation
    process.

    We mention some examples of nonlinear sequence transformations. One
    is  the famous epsilon algorithm \cite{Wynn56a} 
    It computes the upper half of
    the Pad\'e table for the power series $f(\beta)$ according to
    \cite{Shanks55,Wynn56a}
    \begin{equation}
    \epsilon_{2 k}^{(n)}\; = \; [ \, n \, + \, k \, / \, k \, ] \, ,
    \qquad (k\ge 0\;,n\ge 0)\>.
    \end{equation}
    Additionally, the epsilon algorithm is related to the Shanks
    transformation \cite{Shanks55}.
    Unfortunately, the epsilon algorithm is unable to sum several
    important perturbation series, for instance the RSPT series for the
    ground state of the octic anharmonic oscillator \cite{Weniger94}.

    As a second nonlinear sequence transformation, we mention the  \S\
    transformation of Weniger \cite[Sec. 8]{Weniger89} 
that may be
    defined by the recursive scheme
    \begin{subequation}\label{eqJtransRecB}
      \begin{eqnarray}
           \widetilde{D}\di{n}{0} &=& 1/\omega_n\>, \\
           \widetilde{N}\di{n}{0} &=& s_n/\omega_n\>, \\
           \widetilde{D}\di{n}{k} &=&
            \widetilde{D}\di{n+1}{k-1}
               -\Psi\di{n}{k-1}  \widetilde{D}\di{n}{k-1}\>,\qquad
               (k\ge 1)\>, \\
           \widetilde{N}\di{n}{k}  &=&
           \widetilde{N}\di{n+1}{k-1}
              -\Psi\di{n}{k-1}   \widetilde{N}\di{n}{k-1}\>,\qquad
               (k\ge 1)\>, \\
            \frac{\widetilde{N}\di{n}{k}}{\widetilde{D}\di{n}{k}} &=&
           \S\di{n}{k}(\zeta,\{s_n\},\{\omega_n\})
         \end{eqnarray}
    \end{subequation}
    with \cite[Sec. 8.3]{Weniger89}
    \begin{equation} \label{eqPsiWenigerS}
       \Psi\di{n}{k} =
       \frac{(\zeta+n+k)(\zeta+n+k-1)}{(\zeta+n+2k)(\zeta+n+2k-1)}\>
       \qquad (\zeta>0)\>.
    \end{equation}
    The \S\ transformation depends on an auxiliary sequence
    $\{\omega_n\}_{n=0}^{\infty}$ of {\it remainder estimates} with
    $\omega_n\ne 0$. Several variants for the choice of the latter may
    be considered \cite{HomeierWeniger95,Homeier96Hab,Homeier99sls}.
    For the choice $\omega_n= s_{n+1}-s_n$, originally proposed by
    Smith and Ford \cite{SmithFord79}, one obtains the $\tilde t$
    variant (or $d$ variant in the notation of Weniger
    \cite{Weniger89}) , i.e.,
    \begin{equation}
      {}^{\tilde t}\S\di{n}{k}(\zeta,\{s_n\}) =
      \S\di{n}{k}(\zeta,\{s_n\},\{s_{n+1}-s_n\})\>.
    \end{equation}
    Eq.\ (\ref{eqJtransRecB}) is essentially one of the recursive
    schemes for the computation of the \J\ transformation which is a
    rather general and well-studied sequence transformation that covers
    many of the most successful transformations as special cases
    \cite{Homeier93,Homeier94ahc,Homeier95,Homeier96aan,%
    Homeier96Hab,Homeier98ots,Homeier99sls}. The particular choice
    (\ref{eqPsiWenigerS}) corresponds to the fact that the \S\
    transformation is identical to a special case of the \J\
    transformation, namely the case $p=3$ of the ${}_p\mathbf{J}$
    transformation \cite{Homeier94ahc,Homeier96aan,Homeier99sls}.

    Numerical comparison of results of the summation of the RSPT series for the
    ground state of the quartic AHO using the epsilon algorithm and the
    $\tilde t$-variant of the \S\ transformation shows that the
    \S\ transformation is more powerful than the epsilon algorithm for
    the summation of this particular perturbation series  that is
    rather strongly divergent. \cite{Weniger94}

    An important additional consideration stems from regarding the
    perturbation series as function for complex $\beta$, i.e., one
    considers an analytic continuation of the formal power series to a
    function in the complex plane. Then, in principle, one has to allow
    for several branches of such a function. This question was studied
    intensively for the quartic AHO by Bender and Wu
    \cite{BenderWu69,BenderWu71,BenderWu73} and Simon
    \cite{Simon70,Simon72}.
    $E(\beta)$ has a third-order branch point at
    $\beta=0$ that dominates the behavior for large $\beta$, i.e.,
    $E(\beta)=O(\beta^{1/3})$ for $\beta\to\infty$.
    In the case of AHOs, often renormalized
    series are used that correspond --- up to a factor --- to a
    reexpansion of the original series in a transformed variable
    $\kappa$. The relation between $\kappa$ and $\beta$ for the quartic AHO
    may be obtained using Symanzik scaling 
    $x\to (1-\kappa)^{1/4}x$ plus a certain minimization criterion to be
    \cite{VinetteCizek89,WenigerCizekVinette93}
    \begin{equation}
      \beta = \frac{\kappa}{3(1-\kappa)^{3/2}}\>.
    \end{equation}
    After renormalization, one has to consider a Hamiltonian $\widehat
    H(\kappa)=\widehat H_0 + \kappa \widehat V$ with an unperturbed Hamiltonian $\widehat
    H_0=-d^2/dx^2 + x^2$ and a perturbation $\widehat V=(x^4-3x^2)/3$.
    Further, the eigenvalues of $\widehat H(\kappa)$ and $H=H(\beta)$ are
    related by $\widehat E(\kappa)=(1-k)^{1/2} E(\beta)$.
    Besides making the renormalized expansion less divergent,
    the effect of such a renormalization is that the interval
    $(0,\infty)$ for $\beta$ corresponds to a finite interval, e.g., to
    $(0,1)$ for $\kappa$.

    Thus, $E(\beta)$ is a multiple-valued function for complex
    $\beta$ \cite{Simon70,Simon72}. This type of analytic structure is
    not easy to simulate using rational approximants in $\beta$ that
    result on applying either the epsilon algorithm or the \S\
    transformation (or other Levin-type sequence transformations
    \cite{Homeier99sls}) to the RSPT series in $\beta$ of $E(1,\beta)$.
    Such a behavior of $E(1,\beta)$ can  much easier be approximated by
    algebraic functions in $\beta$ instead of rational functions. These
    algebraic functions are nothing but zeroes of polynomials with
    coefficients that depend on $\beta$. For AHOs, this seemingly first
    was recognized by Cizek and coworkers who considered polynomials
    of a special structure, the so-called \emph{effective
    characteristic polynomials} and applied these to sum the divergent
    AHO RSPT series \cite{CizekWenigerBrackenSpirko96}. Effective
    characteristic polynomials have also been applied successfully for
    the extrapolation of quantum chemical many-body perturbation series
    \cite{Homeier96cee} and been proven to be size-extensive, i.e., to
    scale correctly with the particle number \cite{Homeier97tse} in
    this context. Later, Sergeev
    and Goodson \cite{SergeevGoodson98soa} used more general
    polynomials that give rise to \emph{algebraic approximants}
    \cite[Sec.\ 8.6]{BakerGravesMorris96} for the summation of the ground
    state of AHOs. Algebraic approximants are generalizations of the quadratic
    approximants introduced by Shafer \cite{Shafer74}. These 
    types of algebraic approximations will be discussed more thoroughly
    in later sections.

    All the methods mentioned above and most other methods for the
    summation of divergent perturbation series only use the partial
    sums or the coefficients  of a single series, i.e., the RSPT series
    of one eigenvalue for a particular value of the coupling constant
    $\beta$. The rest of the methods may be classified as
    \emph{multipoint} or \emph{multiseries} methods.

    Multipoint methods are those where still a single eigenvalue
    $E^{(I)}(\beta)$ is approximated but several Taylor expansions at
    different points $\beta_i$ are combined to construct the
    approximation. The most prominent example for such methods are
    multipoint Pad\'e approximants \cite{BakerGravesMorris96}. It
    should be noted that the large-coupling regime $\beta\to\infty$
    corresponds to $\kappa\to 1$ in the renormalized treatment and can
    be included in a multipoint treatment. For instance, Cizek and
    coworkers used two-point Pad\'e approximants for this purpose
    \cite{CizekWenigerBrackenSpirko96}. For details of such an
    approach, the reader is referred to the literature, e.g. Refs.\
    \cite{CizekWenigerBrackenSpirko96,Znojil93}.

    In the present work, we consider multiseries methods. These are
    different from multipoint methods since the aim is to approximate
    several functions depending on $\beta$ simultaneously using their
    Taylor series at a common point, e.g., $\beta=0$. This multiseries
    problem occurs naturally in quantum mechanical eigenvalue problems
    since the Hamilton operator normally possesses several discrete
    eigenvalues. Using the same splitting of the Hamiltonian in
    unperturbed operator and perturbation involving the same coupling
    constant $\beta$ for all these eigenvalues considered as functions
    of $\beta$, it is clear that the RSPT
    series of the form (\ref{eqEIbeta}) for several values of $I$ are
    not independent. In the context of rational approximation, this multiseries problem 
    leads
    to simultaneous Pad\'e approximants \cite[Sec. 8.1]{BakerGravesMorris96} or vector 
    Pad\'e approximants \cite[Sec. 8.4]{BakerGravesMorris96}. On
    the other hand, combining the idea of a multiseries method and the
    approximation by algebraic functions, it is rather natural to
    construct a polynomial with coefficients that depend on $\beta$ by
    using the coefficients of several perturbation series
    simultaneously \cite{Homeier96Hab}. The resulting algebraic
    approximations are called \emph{polynomial-type multiseries approximants} and are
    the main topic of the present article. These are related to, but
different from Hermite-Pad\'e approximants as is explained later.

    The possible combination of the multipoint and the multiseries
    concepts leading  to the construction of multipoint multiseries
    approximants is not considered in the present work.

    The outline of the article is as follows. First, we will discuss
    the relation between effective characteristic polynomials and
    algebraic approximants. Then, multiseries approximants are
    introduced and several special types of these approximants are
    defined. Some properties of multiseries approximants are derived.
    In the last section, numerical test results for the case of the
    unrenormalized quartic AHO are presented.

\section{Effective Characteristic Polynomials}

   In this section, we sketch effective characteristic polynomials and
some of their properties that
are known in the literature, and point out that they are special cases
of algebraic approximants. Also, the aim is to motivate the extension to
a multiseries setting discussed in later sections.

   Consider a polynomial of degree $N$ in $E$ with coefficients that
   are polynomials in $\beta$ of the form
   \begin{equation}
     P_N(E)
    =
    \sum_{j=0}^{N} E^j \,\sum_{k=0}^{N-j} f_{N,j,k} \beta^k\>,\qquad
    f_{N,N,0}=1\>.
   \end{equation}
   Such a polynomial depends on $(N+3)N/2$ coefficients $f_{N,j,k}$ and
   is called an \emph{effective characteristic polynomial}
   \cite{CizekWenigerBrackenSpirko96,Homeier96cee,Homeier97tse}. The
   reason for the nomenclature is that characteristic polynomials of
   this \emph{form} arise in the linear variational method in  an
   orthonormal basis $\phi_j$, $j=1,\dots,N$ for Hamiltonians
   $H=H_0+\beta V$, and usually, the $f$'s are computed via matrix
   elements of $H$ in this basis. For \emph{effective} characteristic
   polynomials, however, the $f$'s are obtained from a perturbation
   series
   \begin{equation}\label{eqEinfty}
     E(\beta) = \sum_{j=0}^{\infty} E_j \beta^j
   \end{equation}
   by requiring that
   \begin{equation}
     P_N(E(\beta)) = O(\beta^{N(N+3)/2})
   \end{equation}
   holds for $\beta\to 0$. This leads to a system of linear equations for the $f$'s 
   with as many equations as there are unknowns. If
   this system possesses a solution, the $f$'s and thus, the effective
   characteristic polynomials are uniquely defined. For known $f$'s,
   the eigenvalues  are approximated by zeroes of $P_N$ that are
   denoted by $\Pi_{N,j}$, $j=1,\dots,N$ and are called effective
   characteric polynomial approximants. The branch that reproduces the
   input data is simply denoted by $\Pi_N$ and called the physical
   branch. In order to discuss the dependence of the approximants on
   the coefficients $E_j$ of the series (\ref{eqEinfty}), the explicit
   notation $\Pi_{N,j}[E_0,\dots,E_M]$ is used where $M=N(N+3)/2-1$.

   It should be noted that the physical branch can become complex.
   Then, for $N=2$ both branches yield  complex results, for instance.
   This, of course, is not reasonable for the computation of discrete
   eigenvalues of the Hamiltonian that is a self-adjoint operator, and
   indicates a breakdown of the perturbative approach. Complex values
   of the approximants, however, make sense for the description of
   resonances or tunneling processes \cite[Chap. 8]{Adams94},\cite{Kleinert93b}, and thus, there are
   problems where the possibility of obtaining complex approximations
   from real perturbation series is a desired feature.

We now discuss some properties of these approximants. Note that equivalence and invariance properties of
   algebraic approximants are known from the literature \cite[Sec. 8.6]{BakerGravesMorris96} and 
this is relevant since the latter approximants are generalizations of effective characteristic polynomial approximants as discussed below.

   As shown independently in the literature \cite{Homeier96cee,Homeier97tse} before the connection to
algebraic approximants was recognized, the
   effective characteristic polynomial approximant $\Pi_2$ is 
invariant
   under a repartioning of the Hamiltonian where $H_0$ is replaced by
   the new unperturbed operator $(1-\alpha) H_0$ and $\alpha$ is some
   constant. Denoting the corresponding coefficients of the RSPT
   eigenvalue series by $E_j(\alpha)$, one has
   \begin{equation}
     \Pi_2[E_0,\dots,E_4] = \Pi_2[E_0(\alpha),\dots,E_4(\alpha)]\>.
   \end{equation}
   This means that the approximant is invariant under the repartioning
   as the true eigenvalue has to be. 
   In Ref.\ \cite{Homeier97tse} it is shown that a particular effective
   characteristic polynomial approximant $\Pi_{N,j}$ for some $N$ and
   $j$ with $1\le j\le N$ has the scaling property
   \begin{equation}
     \Pi_{N,j}[c\,E_0,\dots,c\,E_{M}] = c\,\Pi_{N,j}[E_0,\dots,E_{M}]\>.
   \end{equation}
   This is important to guarantee size-extensitivity, i.e., correct
   scaling with particle number in many-body perturbation theory
   \cite{Homeier97tse}.

   We now show that effective characteristic polynomial approximants
   are special \emph{algebraic approximants}.
   These are generalizations of Pad\'e approximants. 
   Algebraic approximants are constructed via polynomials
   $A^{(k)}$ in $\beta$ with ${\rm deg} (A^{(k)})=d_k$, $k=0,\dots,N$
   such that for $Q=-1+N+\sum_{k=0}^N d_k$ and for a given power series
   $E(\beta)$ as defined in Eq.\ (\ref{eqEinfty}) the relations
   \begin{equation}
     P_N(E(\beta))=\sum_{k=0}^{N} A^{(k)}(\beta) [E(\beta)]^k = O(\beta^{Q+1})
   \end{equation}
   are satisfied for $\beta\to 0$. Since this defines the  polynomials
   only up to a common factor, we additionally demand $A^{(N)}(0)=1$ as
   a normalization. For given polynomials $A^{(k)}$, the
   algebraic approximants $E_{[d_0,d_1,\dots,d_N],j}$ are defined as the zeroes of the polynomial
   $P_N$ according to
   \begin{equation}
     P_N\left(E_{[d_0,d_1,\dots,d_N],j}\right) =0\>, \qquad
     (j=1,\dots,N)\>,
   \end{equation}
   and depend on $Q+1$ coefficients $E_0,\dots,E_Q$ that play the role
   of the input data. Again, one may define the physical branch as that
   one which reproduces the input data via a Taylor expansion in
   $\beta$.

   Comparison with the definition of effective characteristic
   polynomial approximants reveals that the latter are nothing but the
   special algebraic approximants
   \begin{equation}
     \Pi_{N,j} = E_{[N,N-1,N-2,\dots,0],j}\>, \qquad (j=1,\dots,N)\>.
   \end{equation}

   It should be noted that algebraic approximants
\begin{itemize}
\item are special cases of Hermite-Pad\'e approximants and are
described by a quite elaborate mathematical theory
\cite[Sec.\ 8.5, 8.6]{BakerGravesMorris96}
\item are useful for multi-valued functions (analytic continuation to another
Riemann sheet)
\item have been applied successfully to anharmonic oscillators
\cite{SergeevGoodson98soa} as noted in the introduction.
\end{itemize}

\section{Polynomial-type Multiseries Approximants}
In this section, we define the polynomial-type multiseries approximants,
and introduce three special cases ($\Pi$-, D-, and P-type approximants).
The relation to Hermite-Pad\'e approximants is pointed out. An example
is given for $\Pi$-type approximants. Further, it is proved that the
polynomial coefficients of P-type approximants can be computed
recursively.

     As indicated in the introduction, the key idea of the present work
     is to use coefficients of several perturbation series for
     different eigenvalues to compute the algebraic approximants.

     For the definition of \emph{polynomial-type multiseries approximants}\footnote{
These are called simply multiseries approximants in the following.}, we
consider a polynomial
   \begin{equation}\label{eqPoly}
     P_N(z)=\sum_{k=0}^{N} A^{(k)}(\beta) z^k 
   \end{equation}
where the coefficients $A^{(k)}$ are polynomials
     in $\beta$ with ${\rm deg} (A^{(k)})=d_k\ge 0$ for
$k=1,\dots, N$.
     They satisfy for $\beta \to 0$ the order conditions
     \begin{equation}\label{eqOrderMult}
       P_N( E^{(I)}(\beta)) = O(\beta^{Q_I+1}) \>, \qquad (I=1,\dots, S)
     \end{equation}
     for $S$ given series $E^{(I)}(\beta)$ of the form
     (\ref{eqEIbeta}) and $S$ parameters $Q_I$ restricted by demanding
     \begin{equation}\label{eqConsistency}
       \sum_{I=1}^{S} (Q_I+1)=N+\sum_{k=0}^N d_k\>.
     \end{equation}
     The normalization condition is $A^{(N)}(0)=1$. The latter and
     the order conditions (\ref{eqOrderMult}) again lead to a system of
     linear equations for the coefficients of the polynomials $A^{(k)}$
     with as many unknowns as equations if (\ref{eqConsistency}) is
     satisfied. Hence the polynomials $A^{(k)}$ are uniquely defined if
     the linear system possesses a solution as is assumed in the
     following.

     Given the polynomials $A^{(k)}$, the multiseries approximants\newline
     $E^{[Q_1,\dots,Q_S]}_{[d_0,d_1,\dots,d_N],j}$ are
     defined as the zeroes of $P_N$, i.e.,
     \begin{equation}
       P_N\left(E^{[Q_1,\dots,Q_S]}_{[d_0,d_1,\dots,d_N],j}\right)
       =0 \>.
     \end{equation}
     These approximants depend on $\beta$ and the coefficients
     \begin{equation}
       E^{(I)}_0,\dots,E^{(I)}_{Q_I}
       \>,\qquad I=1,\dots, S\>.
     \end{equation}

We are especially interested in the case that $S=N$ which means that as
many perturbation series are used as there are roots of $P_N$, i.e., as
there are approximants. In this case, the consistency condition becomes
     \begin{equation}\label{eqConsistency1}
       \sum_{I=1}^{N} Q_I=\sum_{k=0}^N d_k\>.
     \end{equation}
Therefore, we define the following special cases:

\begin{description}
   \item[$\Pi$-type approximants with $S=N$]
     \begin{equation}
       \Pi_{[N],j} = E^{[N,N-1,N-2,\dots,1]}_{[N,N-1,N-2,\dots,1,0],j}\>, \qquad
     (j=1,\dots,N)\>.
     \end{equation}
     Here, Eq.\ (\ref{eqConsistency1}) is obviously satisfied.
   \item[$D$-type approximants with $S=N$ for even $N=2D$]
     \begin{equation}
       D_{[N],j} = E^{[N,N-1,N-2,\dots,1]}_{[D,\dots,D],j}\>, \qquad
     (j=1,\dots,N)\>.
     \end{equation}
     Here, Eq.\ (\ref{eqConsistency1}) is  satisfied: The left hand
     side is $N(N+1)/2$, the right hand side is $D(N+1)$, and both
     agree since $N=2D$.
   \item[$P$-type approximants with $S=N$]
     \begin{equation}
       P_{[Q],j} = E^{[Q,Q,\dots,Q]}_{[Q,Q,\dots,Q,0],j}\>, \qquad
     (j=1,\dots,N)\>.
     \end{equation}
     Here again, Eq.\ (\ref{eqConsistency1}) is obviously satisfied.
\end{description}
$\Pi$-type approximants were introduced by the author
\cite{Homeier96Hab}, D- and P-type approximants are new.

We note that these multiseries approximants
\begin{itemize}
\item are related to, but \emph{different} from Hermite-Pad\'e approximants \cite[Sec.\ 8.5]{BakerGravesMorris96},
To be more specific, we remark that the order conditions (\ref{eqOrderMult}) are different from
those satisfied by the Hermite-Pad\'e polynomials $a_{I,m_I}(\beta)$ 
that are contructed from the same $S$ power series
$E^{(I)}(\beta)$, $I=1,\dots,S$ via, e.g.,   
\begin{equation}
  a_{I,0}(\beta) + \sum_{I=1}^{S} \sum_{m_I=1}^{M_I} a_{I,m_I}(\beta)
[E^{(I)}(\beta)]^{m_I} = O(\beta^\tau)
\end{equation}
for suitable $M_I$ and $\tau$. This should be compared to Eqs.\
(\ref{eqPoly}) and (\ref{eqOrderMult}).
\item are useful for multi-valued functions (analytic continuation to another
Riemann sheet),
\item have no convergence theory so far, and
\item are possibly cheaper than algebraic approximants since the numerical effort
to compute several short perturbation expansions is much less than one long
expansion.
\end{itemize}

As a simple example, we consider the harmonic oscillator problem with
Hamiltonian
\begin{equation}
H = -\frac{d^2}{dx^2 } + (1+\beta) x^2\>, \quad H_0=H-\beta x^2
\end{equation}
with eigenvalues
\begin{equation}
E^{(I)} = (2 I - 1 ) \sqrt{1+\beta}\>.
\end{equation}
The coefficients of the perturbation series follow by Taylor expansion
of the $E^{(I)}$:
\begin{subequation}\label{eqHOtrus}
\begin{eqnarray}
 {{E}^{(1)}}  &=& \underline{1 + { \frac {1}{2}}\,{ \beta } -
{ \frac {1}{8}}\,{ \beta }^{2}} + {
\frac {1}{16}}\,{ \beta }^{3} - { \frac {5}{128}}\,
{ \beta }^{4} + ... \\
 {{E}^{(2)}} &=& \underline{3 + { \frac {3}{2}}\,{ \beta }} -
{ \frac {3}{8}}\,{ \beta }^{2} + {
\frac {3}{16}}\,{ \beta }^{3} - { \frac {15}{128}}
\,{ \beta }^{4} + ...
\end{eqnarray}
\end{subequation}
Then,  using the series for $I=1$ up to $E^{(1)}_2$ and
$I=2$ up to $E^{(1)}_1$ as indicated by underlines in Eq.\
(\ref{eqHOtrus}), we obtain the following
results for $\Pi_{[2]}$:
\begin{subequation}
\begin{eqnarray}
  {\Pi_{[2],1}}&=&2 + {
\beta } - { \frac {1}{2}}\,\sqrt {4 + 4\,{ \beta }
 + 2\,{ \beta }^{2}}\>,  \nonumber\\
&=& \underline{1 + { \frac {1}{2}}\,{ \beta } -
{ \frac {1}{8}}\,{ \beta }^{2} + {
\frac {1}{16}}\,{ \beta }^{3}} - { \frac {3}{128}}\,
{ \beta }^{4} + ... \\
  {\Pi_{[2],2}}&=&2 + { \beta } + { \frac {1}{2}}\,
\sqrt {4 + 4\,{ \beta } + 2\,{ \beta }^{2}}
\nonumber \\
&=& \underline{3 + { \frac {3}{2}}\,{ \beta }} +
{ \frac {1}{8}}\,{ \beta }^{2} - {
\frac {1}{16}}\,{ \beta }^{3} + { \frac {3}{128}}\,
{ \beta }^{4} + ...
\end{eqnarray}
\end{subequation}
Here, we have underlined the terms that are correct in comparison with
the exact results.
Thus, we gain one order for $E^{(1)}$ in comparison to the input data
while there is no gain in $E^{(2)}$.

The following theorem shows that the coefficients in $P$-type
polynomials leading to approximants $P_{[Q+1]}$ satisfy the equations
for the polynomials leading to approximants $P_{[Q]}$, and hence, these
polynomials can be computed recursively.

\begin{theorem} Define
\begin{equation}
P_N^{(Q)}(E) = E^N + \sum_{j=0}^{N-1} E^j \sum_{k=0}^{Q} a_k^{(j,Q)} \beta^k
\end{equation}
for all $Q$ and given $N$. The zeroes of the polynomial $P_N^{(Q)}$ are the
$P$-type approximants $P_{[Q]}$, if for $\beta\to 0$
\begin{equation}
P_N^{(Q)}\left(\sum_{\ell=0}^{Q} E^{(I)}_\ell \beta^{\ell}\right) =
O(\beta^{Q+1})\>, \quad (I=1,\dots,N)
\end{equation}
holds. Assume that for $I=1,\dots,N$
\begin{equation}
P_N^{(Q+1)}\left(\sum_{\ell=0}^{Q+1} E^{(I)}_\ell \beta^{\ell}\right) = O(\beta^{Q+2})
\end{equation}
holds for $\beta\to 0$ whence the zeroes of this polynomial are the
$P$-type approximants $P_{[Q+1]}$.
Define
\begin{equation}
\widetilde P_N^{(Q)}(E) = E^N + \sum_{j=0}^{N-1} E^j \sum_{k=0}^{Q}
a_k^{(j,Q+1)} \beta^k\>.
\end{equation}
Then for $\beta\to 0$,
the relations
\begin{equation}
\widetilde P^{(Q)}\left(\sum_{\ell=0}^{Q} E^{(I)}_\ell \beta^{\ell}\right) = O(\beta^{Q+1})
\end{equation}
hold for $I=1,\dots,N$, and thus, $\widetilde P_N^{(Q)}=P_N^{(Q)}$.
\end{theorem}
\begin{pf}
Put
\begin{equation}
E^{(I,Q)}(\beta)=\sum_{\ell=0}^{Q} E^{(I)}_\ell \beta^{\ell}\>.
\end{equation}
Then direct calculation shows
\begin{eqnarray}
\widetilde P^{(Q)}(E^{(I,Q)}(\beta)) &=&
 -\beta^{Q+1} [E^{(I,Q)}(\beta)]^{N-1} N E^{(I)}_{Q+1} \nonumber\\
& & - \beta^{Q+1} \sum_{j=0}^{N-1} j  [E^{(I,Q)}(\beta)]^{j-1}
E^{(I)}_{Q+1}
\sum_{k=0}^{Q} a_k^{(j,Q+1)} \beta^k\nonumber \\
& & - \beta^{Q+1} \sum_{j=0}^{N-1} [E^{(I,Q)}(\beta)]^{j}
a_{Q+1}^{(j,Q+1)} \nonumber\\
& & + O(\beta^{Q+2})\nonumber \\
&=&O(\beta^{Q+1})
\end{eqnarray}
\qed
\end{pf}

Recursive algorithms are also known for certain algebraic approximants
\cite{SergeevGoodson98soa}, and, more generally for the computation of Hermite-Pad\'e
polynomials \cite[Sec. 8.5]{BakerGravesMorris96}.

\section{Numerical Tests}

In the test cases, we always treat the quartic AHO with Hamiltonian as
defined in Eq. (\ref{eqquarticAHO}) for small $\beta$
without any renormalization. The coefficients in all the perturbation
series used below have been computed using RSPT up to order 20 in the
wave functions, followed by Taylor expansion of the Rayleigh-Ritz
expectation value with this wave function leading to the coefficients
in the perturbation series for the energies up to order 41. The
computations were done in \maple.

\begin{table}
\caption{Quadratic P-type Approximants for the
quartic AHO with $\beta=1/100$}\label{tabP2}
\begin{tabular}{rll}
\hline
$Q$ & $P_{[Q],1}$ & $P_{[Q],2}$ \\
\hline
            2 & 1.007375         & 3.03646 \\
            3 & 1.0073736        & 3.03653 \\
            4 & 1.00737368       & 3.036525 \\
            5 & 1.007373671      & 3.0365254 \\
            6 & 1.0073736722     & 3.03652530 \\
            7 & 1.00737367206    & 3.036525306 \\
            8 & 1.00737367208    & 3.0365253043 \\
            9 & 1.007373672081   & 3.0365253045 \\
           10 & 1.0073736720815  & 3.03652530451 \\
           11 & 1.00737367208137 & 3.036525304514 \\
           12 & 1.00737367208139 & 3.0365253045131 \\
           13 & 1.00737367208138 & 3.0365253045134 \\
           14 & 1.00737367208138 & 3.03652530451334 \\
           15 & 1.00737367208138 & 3.03652530451335 \\
\hline
           $\infty$ & 1.00737367208138 & 3.03652530451335 \\
\hline
 \end{tabular}
\end{table}
\begin{table}\caption{ $\Pi$-type Approximants for the
quartic AHO with $\beta=1/100$}\label{tabPiN}
\begin{tabular}{rll}
\hline
$N$ & $\Pi_{[N],1}$ & $\Pi_{[N],2}$ \\
\hline
2 &   1.007371         &  3.0376 \\
3 &   1.0073738        &  3.03650 \\
4 &   1.007373667      &  3.0365266 \\
5 &   1.0073736724     &  3.03652522 \\
6 &   1.00737367206    &  3.036525310 \\
7 &   1.00737367208    &  3.0365253040 \\
8 &   1.0073736720812  &  3.03652530456 \\
9 &   1.00737367208140 &  3.036525304509 \\
10 &  1.00737367208138 &  3.0365253045138 \\
\hline
$\infty$ & 1.00737367208138 & 3.03652530451335 \\
\hline
 \end{tabular}\\
\end{table}
\begin{table}
\caption{ $D$-type Approximants for the
quartic AHO with $\beta=2/10$}\label{tabDN}
\begin{tabular}{rll}
\hline
$N$ & $D_{[N],1}$ & $D_{[N],2}$ \\
\hline
2     & 1.11     & 3.8 \\
4     & 1.117    & 3.53 \\
6     & 1.1181   & 3.534 \\
8     & 1.11826  & 3.5377 \\
10    & 1.11828  & 3.5386 \\
12    & 1.118291 & 3.5389 \\
\hline
Exact & 1.118293 & 3.5390 \\
\hline
 \end{tabular}
\end{table}

In Table \ref{tabP2}, we consider quadratic P-type approximants
($N=S=2$) for various values of $Q$. The calculation of approximants
$P_{[Q],j}$ requires two $Q$-th order perturbation series for the two
lowest eigenvalues. The results show that both eigenvalues are well
approximated for $Q=15$. The convergence of the approximants is
somewhat faster for the lower eigenvalue.

In Table \ref{tabPiN}, we consider  $\Pi$-type approximants
for various values of $N$. The calculation of approximants
$\Pi_{[N],j}$ requires $N$ perturbation series with orders $ N,N-1,\dots,1$.\\
Only the results for the two
lowest eigenvalues are displayed for comparison reasons.
The data show that $\Pi$-type approximants converge somewhat faster
than $P$-type approximants in the example treated.
The results show that both eigenvalues are well
approximated for $N=10$. The convergence of the approximants is
somewhat faster for the lower eigenvalue.

For large $\beta$, both $\Pi$- and $P$-type approximants break down
and start to produce complex approximants.

In Table \ref{tabDN}, we consider  $D$-type approximants
for various values of $N$ for a larger $\beta$.
The calculation of approximants
$D_{[N],j}$ requires $N$ perturbation series with orders $
N,N-1,\dots,1$ (even $N$).
Again, only the results for the two
lowest eigenvalues are displayed for comparison reasons.
The results indicate that this type of multiseries approximant
can be useful for somewhat larger $\beta$. Comparison values are taken from
Ref.\ \cite{Meissner95}.

In further studies it is
planned to study these approximants also for renormalized perturbation
series in order to see whether for these the range of applicability of
the approximants is extended as is the case for several other methods.

\section{Summary}
\begin{itemize}
\item Effective characteristic polynomial approximants are special
      algebraic approximants.
\item Polynomial-type multiseries approximants combine the information of several
      perturbation series and allow the summation of divergent series.
      They are different from Hermite-Pad\'e approximants.
\item Perturbation series for these multiseries approximants are less costly
      to calculate than for algebraic approximants
      (several short series vs. one long expansion).
\item The simultaneous calculation of several eigenvalues is possible.
      Higher eigenvalues converge slower.
\item The large number of variants of multiseries approximants should
      be explored further.
\end{itemize}

\begin{acknowledgements}
{The author thanks the organizers of ICRA99 for their
superb organizational work and hospitality and for the opportunity to
present the material given above as talk at this pleasant conference.
For stimulating discussions regarding effective characteristic polynomials,
the author is thankful to Prof. Dr. J. {\v C}{\'\i}{\v z}ek and Priv.-Doz.\ Dr.\ E.\ J.\
Weniger. The financial support of the \emph{Deutsche
Forschungsgemeinschaft}, the \emph{Fonds der Chemischen Industrie}, and
the \emph{Verein der Freunde der Universit\"at Regensburg} is gratefully
acknowledged.}
\end{acknowledgements}


\begin{thebibliography}{00}

\bibitem{Adams94}
Adams, B.~G.: 1994, {\em Algebraic Approach to Simple Quantum Systems}.
\newblock Berlin: Springer.

\bibitem{BakerGravesMorris96}
{Baker, Jr.}, G.~A. and P. Graves-Morris: 1996, {\em {P}ad{\'e} approximants}.
\newblock Cambridge (GB): Cambridge U.P., second edition.

\bibitem{BenderWu69}
{B}ender, C.~M. and T.~T. {W}u: 1969, `Anharmonic oscillator'.
\newblock {\em Phys. Rev.} {\bf 184}, 1231--1260.

\bibitem{BenderWu71}
{B}ender, C.~M. and T.~T. {W}u: 1971, `Large-order behavior of perturbation
  theory'.
\newblock {\em {P}hys. Rev. Lett} {\bf 27}, 461--465.

\bibitem{BenderWu73}
{B}ender, C.~M. and T.~T. {W}u: 1973, `Anharmonic oscillator. {II.} {A} study
  in perturbation theory in large order'.
\newblock {\em {P}hys. Rev. D} {\bf 7}, 1620--1636.

\bibitem{CizekWenigerBrackenSpirko96}
{\v{C}\'{\i}\v{z}ek}, J., E.~J. Weniger, P. Bracken, and V. {\v{S}pirko}: 1996,
  `Effective characteristic polynomials and two-point {P}ad{\'e} approximants
  as summation techniques for the strongly divergent perturbation expansions of
  the ground state energies of anharmonic oscillators'.
\newblock {\em Phys. Rev. E} {\bf 53}, 2925--2939.

\bibitem{Homeier99sls}
Homeier, H. H.~H., `Scalar {Levin}-type sequence transformations'.
\newblock Invited review for {\em J. Comp. Appl. Math}, \submitted.

\bibitem{Homeier93}
Homeier, H. H.~H.: 1993, `Some Applications of Nonlinear Convergence
  Accelerators'.
\newblock {\em Int. J. Quantum Chem.} {\bf 45}, 545--562.

\bibitem{Homeier94ahc}
Homeier, H. H.~H.: 1994, `A hierarchically consistent, iterative sequence
  transformation'.
\newblock {\em Numer. Algo.} {\bf 8}, 47--81.

\bibitem{Homeier95}
Homeier, H. H.~H.: 1995, `Determinantal representations for the {$\mathcal J$}
  transformation'.
\newblock {\em Numer. Math.} {\bf 71}(3), 275--288.

\bibitem{Homeier96aan}
Homeier, H. H.~H.: 1996a, `Analytical and numerical studies of the convergence
  behavior of the {$\mathcal J$} transformation'.
\newblock {\em J. Comput. Appl. Math.} {\bf 69}, 81--112.

\bibitem{Homeier96cee}
Homeier, H. H.~H.: 1996b, `Correlation Energy Estimators based on
  {M{\o}ller}-{Plesset} Perturbation Theory'.
\newblock {\em J. Mol. Struct. (Theochem)} {\bf 366}, 161--171.

\bibitem{Homeier96Hab}
Homeier, H. H.~H.: 1996c, `{Extrapolationsverfahren f\"ur Zahlen-, Vektor- und
  Matrizenfolgen und ihre Anwendung in der Theoretischen und Physikalischen
  Chemie}'.
\newblock Habilitation thesis, Universit\"at Regensburg.

\bibitem{Homeier97tse}
Homeier, H. H.~H.: 1997, `The size-extensivity of correlation energy estimators
  based on effective characteristic polynomials'.
\newblock {\em J. Mol. Struct. (Theochem)} {\bf 419}, 29--31.
\newblock Proceedings of the {3$^{rd}$ Electronic Computational Chemistry
  Conference}.

\bibitem{Homeier98ots}
Homeier, H. H.~H.: 1998, `On the Stability of the {$\mathcal{J}$}
  Transformation'.
\newblock {\em Numer. Algo.} {\bf 17}, 223--239.

\bibitem{HomeierWeniger95}
Homeier, H. H.~H. and E.~J. Weniger: 1995, `On Remainder Estimates for
  {Levin}-type Sequence Transformations'.
\newblock {\em Comput. Phys. Commun.} {\bf 92}, 1--10.

\bibitem{Kleinert93b}
Kleinert, H.: 1993, {\em Pfadintegrale in Quantenmechanik, Statistik und
  Polymerphysik}.
\newblock Mannheim: B. I. Wissenschaftsverlag.

\bibitem{Meissner95}
Mei{\ss}ner, H.: 1995, `Iterative {Bestimmung} der {Elektronenkorrelation} und
  der station{\"a}ren {Zust{\"a}nde} elektronisch angeregter {Molek{\"u}le}
  sowie anharmonischer {Oszillatoren} mit {Hilfe} der verallgemeinerten
  {Bloch}-Gleichung'.
\newblock Doktorarbeit, Universit{\"a}t Regensburg, Germany.

\bibitem{MeissnerSteinborn97ani}
Mei{\ss}ner, H. and E.~O. Steinborn: 1997, `A New Iterative Method for Solving
  the Time-Independent {Schr{\"o}dinger} Equation Based on the Generalized
  {Bloch} Equation. {I}. {Boson} Systems: {The} Quartic Anharmonic Oscillator'.
\newblock {\em Int. J. Quantum Chem.} {\bf 61}, 777--795.

\bibitem{MeissnerSteinborn97qsa}
Meissner, H. and E.~O. Steinborn: 1997, `Quartic, sextic, and octic anharmonic
  oscillators: {Precise} energies of ground state and excited states by an
  iterative method based on the generalized {Bloch} equation'.
\newblock {\em Phys. Rev. A} {\bf 56}, 1189--1200.

\bibitem{SergeevGoodson98soa}
Sergeev, A.~V. and D.~Z. Goodson: 1998, `Summation of asymptotic expansions of
  multiple-valued functions using algebraic approximants: {Application} to
  anharmonic oscillators'.
\newblock {\em J. Phys. A: Math. Gen.} {\bf 31}, 4301--4317.

\bibitem{Shafer74}
Shafer, R.~E.: 1974, `On quadratic approximation'.
\newblock {\em SIAM J. Num. Anal.} {\bf 11}, 447--460.

\bibitem{Shanks55}
Shanks, D.: 1955, `Non-linear transformations of divergent and slowly
  convergent sequences'.
\newblock {\em J. Math. and Phys. (Cambridge, Mass.)} {\bf 34}, 1--42.

\bibitem{Simon70}
Simon, B.: 1970, `Coupling constant analyticity for the anharmonic oscillator'.
\newblock {\em Ann. Phys. (NY)} {\bf 58}, 76--136.

\bibitem{Simon72}
Simon, B.: 1972, `{T}he anharmonic oscillator: A singular perturbation theory'.
\newblock In: D. Bessis (ed.): {\em Carg{\`e}se lectures in physics}, Vol.~5.
\newblock New York: Gordon and Breach, pp. 383--414.

\bibitem{SmithFord79}
Smith, D.~A. and W.~F. Ford: 1979, `Acceleration of linear and logarithmic
  convergence'.
\newblock {\em SIAM J. Numer. Anal.} {\bf 16}, 223--240.

\bibitem{VinetteCizek89}
Vinette, F. and J. {\v C}{\'\i}{\v z}ek: 1989, `{T}he use of symbolic
  computation in solving some non-relativistic quantum mechanical problems'.
\newblock In: P. Gianni (ed.): {\em Symbolic and Algebraic Computation.
  International Symposium ISSAC '88 -- Rome, Italy}. Berlin, pp. 85--95.

\bibitem{Weniger89}
Weniger, E.~J.: 1989, `Nonlinear sequence transformations for the acceleration
  of convergence and the summation of divergent series'.
\newblock {\em Comput. Phys. Rep.} {\bf 10}, 189--371.

\bibitem{Weniger94}
Weniger, E.~J.: 1994, `{Verallgemeinerte Summationsprozesse als numerische
  Hilfsmittel f\"ur quantenmechanische und quantenchemische Rechnungen}'.
\newblock Habilitationsschrift, Universit{\"a}t Regensburg.

\bibitem{WenigerCizekVinette93}
Weniger, E.~J., J. {\v C}{\'\i}{\v z}ek, and F. Vinette: 1993, `{T}he summation
  of the ordinary and renormalized perturbation series for the ground state
  energy of the quartic, sextic and octic anharmonic oscillators using
  nonlinear sequence transformations'.
\newblock {\em J. Math. Phys.} {\bf 34}, 571--609.

\bibitem{Wynn56a}
Wynn, P.: 1956, `On a device for computing the {$e_m (S_n)$} transformation'.
\newblock {\em Math. Tables Aids Comput.} {\bf 10}, 91--96.

\bibitem{Znojil93}
Znojil, M.: 1993, `{T}he three-point {Pad{\'e}} resummation of perturbation
  series for anharmonic oscillators'.
\newblock {\em Phys. Lett. A} {\bf 177}, 111--120.

\end{thebibliography}

\end{document}